\documentclass[]{tPHM2e}

\usepackage{mathrsfs}
\usepackage{epsfig,bm,dcolumn}
\usepackage{graphicx}
\usepackage{color}
\usepackage{amsmath}

\begin{document}
\doi{10.1080/14786435.20xx.xxxxxx}
\issn{1478-6443}
\issnp{1478-6435}
\jvol{00} \jnum{00} \jyear{2010}

\title{Superconductor-normal metal quantum phase transition in dissipative and non-equilibrium systems}

\author{Fernanda Deus$^{\rm a}$
and Mucio A. Continentino$^{\rm a}$$^{\ast}$\thanks{$^\ast$Corresponding author. Email: mucio@cbpf.br} 
\\\vspace{6pt}  
$^{\rm a}${\em{Centro Brasileiro de Pesquisas F\'{\i}sicas \\
Rua Dr. Xavier Sigaud, 150, Urca \\ 
22290-180, Rio de Janeiro, RJ, Brazil}}
\\\vspace{6pt}\received{v4.5 released May 2010} }

\maketitle

\begin{abstract}

In physical systems, coupling to the environment gives rise to dissipation and decoherence.
For nanoscopic  materials this may be a determining factor of their physical behavior.  However, even for macroscopic many-body systems, if the strength of this coupling  is sufficiently strong, their ground state properties and phase diagram may be severely modified. Also dissipation is essential to allow a system in the presence of 
a time dependent perturbation  to attain a steady,  time independent  state. In this case, the non-equilibrium phase diagram   depends on the intensity of the perturbation and on the strength of the coupling of the system to the outside world. 
In this paper, we investigate the effects of both, dissipation and time dependent external sources in the phase diagram of a many-body system at zero and finite temperatures. For concreteness we consider the specific case of a superconducting layer under the action of an electric field and coupled to a metallic substrate. The former arises from a time dependent vector potential minimally coupled to the electrons in the  layer. We introduce a Keldysh approach that allows to obtain the time dependence of the superconducting order parameter in an adiabatic regime.  
We study the phase diagram of this system as a function of the electric field, the coupling to the metallic substrate and  temperature.  
 \end{abstract}

\section{Introduction}

Recently, with the advent of nanotechnology great interest and effort have been devoted to the study of  decoherence in nano-systems that arises from their coupling to the environment~\cite{ventra}. Due to their small sizes and consequently reduced number of degrees of freedom this is a fundamental feature which is going to determine their properties and eventual applications. Equally important is to understand the behavior of these nano materials under the action of external time dependent perturbations. 

It turns out however that even in macroscopic many-body systems, the effects of dissipation and time dependent perturbations are extremely interesting and important to investigate. These perturbations can radically modify the phase diagram of macroscopic systems, both at zero and finite temperatures. An important and actual question concerns the nature of the phase transitions in nonequilibrium systems, both classical and quantum.  In the quantum case, due to the entanglement of time and space dimensions, including dissipation is sometimes essential to give a correct description of the quantum critical phenomena. The most common example is that of magnetic quantum phase transitions in metals where  the magnetic order parameter couples dynamically in a dissipative way to low energy particle-hole excitations~\cite{hertz}. 

While the mathematical treatment of dissipation may be involved, sometimes requiring a numerical approach~\cite{chakra}, it can be done with the usual methods of equilibrium problems.  On the other hand, when considering the effect of external time dependent perturbations, beyond the regime of linear response, one has to appeal for more unconventional mathematical approaches to deal with the nonequilibrium situation, even in steady state. In this case there is a breakdown of the fluctuation-dissipation theorem~\cite{tyablikov} due to the explicit time dependence of the parameters in the Hamiltonian describing the system~\cite{kadanoff,keldysh}.

In this paper, we investigate the effects of dissipation and of a time dependent vector potential in the phase diagram of a superconducting material~\cite{phillips,sondhi,mitra}. To be specific, we focus on a system consisting of a superconducting layer under the action of an external electric field that arises from a time dependent vector potential that acts exclusively in the  layer.  The latter is coupled to a metallic substrate that represents a source of dissipation and allows for the perturbed system to attain a nonequilibrium, time independent situation.   
The microscopic mechanism for dissipation is the transfer of electrons between the two systems~\cite{mcmillan}, the same type of coupling which gives rise to the proximity effect.

We use a nonequilibrium Keldysh approach and develop a formalism that allows to obtain the time dependence of the superconducting order parameter, for an arbitrary time dependence of the vector potential, under an adiabatic condition. Our method does not resort to the  time dependent Landau-Guinzburg equation neither to  Boltzmann"s equation. 
In this paper, besides introducing the method~\cite{alexis,dissertacao}, we will be mainly concerned with the phase diagram
of the superconducting layer, as modified by dissipation and the electric field. We find that for sufficiently large dissipation and electric field superconductivity in the layer is destroyed at zero temperature. We fully characterize the quantum critical point (QCP) associated with the dissipation induced superconductor-normal metal transition. Furthermore, we discuss the nature of the transition in the presence of the electric field. We show that in a certain limit this reduces to a \textit{depairing transition}~\cite{bardeen,depairing,depairing2,yu,degennes}  in the presence of dissipation.

The problem considered in this paper is  a classical one in the study of superconductivity from both, experimental and theoretical aspects. It has been treated  in the literature by many authors and  at different times~\cite{mcmillan,bardeen,depairing,depairing2,yu,degennes,maki}, always with renewed interest~\cite{phillips,sondhi,mitra,takei,hogan,feldman}.  As fabrication and experimental techniques improve, new aspects of this problem are revealed requiring more sophisticated theoretical approaches. The present work introduces a novel theoretical treatment and focus on the new aspects of quantum criticality in this paradigmatic system~\cite{livro}.

\section{Hamiltonian}

We consider the following Hamiltonian~\cite{mitra} describing the superconducting layer, the metallic substrate and their coupling,
\begin{eqnarray}\label{hamiltonian}
H = H_{bath} + H_{layer} + H_{bath-layer},
\end{eqnarray}
where $ H_{layer}$ describes the two-dimensional (2D) superconductor whose physical properties we are interested in.  $H_{bath}$ takes into account the reservoir  and  $H_{bath-layer}$ in an obvious notation the coupling between the 2D superconductor and the bulk metal. This coupling represents a source for dissipation. Explicitly these Hamiltonians are given by,
\begin{eqnarray}
H_{layer} &=& \sum _{k, \sigma} \frac{1}{2m}\!\left(\! \frac{\nabla}{i} \! -\! \frac{e}{\hbar c}\vec{A}(t)\!\right) ^{2} d_{k,\sigma}^{\dagger} d_{k,\sigma}\! -\!  \lambda \sum _{k} d_{k,\uparrow}^{\dagger}d_{k,\downarrow}^{\dagger}d_{k,\downarrow}d_{k,\uparrow}; \\
H_{bath} &=& \sum _{k_{z}k\sigma} \epsilon _{k_{z},k,\sigma}^{b} c_{k_{z},k,\sigma}^{\dagger}c_{k_{z},k,\sigma}; \\
H_{bath-layer} &=& \sum _{k_{z}k\sigma} \left( t_z c_{k_{z},k,\sigma}^{\dagger} d_{k, \sigma} + H.c.\right) ,
\label{hbl}
\end{eqnarray}
The vector potential $\vec{A}(t) = -c\vec{E}t$   ($\hbar=1$) gives rise to an electric field $\vec{E}=-1/c(\partial \vec{A}/\partial t)$ that acts only in the superconducting layer. The quantity $\epsilon _{k_{z},k,\sigma}^{b}$ is the kinetic energy of the electrons in the bath (substrate). Notice that  $\vec{k} = (k_{\perp}, k_{z})$ and for simplicity we write $k_{\perp}=k$. The coupling between the layer and substrate [Eq.~(\ref{hbl})] has an intensity $t_z$ and transfers electrons from the layer to the substrate and vice-versa. It does not conserve momentum since the superconductor is strictly 2D.  

Applying a  BCS decoupling~\cite{aproxbcs} to $H_{layer}$,  we get
\begin{eqnarray}
H_{layer} = \sum _{k, \sigma} \epsilon \left( k - \frac{e\vec{A}(t)}{\hbar c}\right) d_{k \sigma}^{\dagger} d_{k \sigma} - \frac{ 1}{2} \sum_{k}  \left(\Delta_{k}   d_{k\uparrow}^{\dagger}d_{-k \downarrow}^{\dagger} + H.c.\right) ,
\end{eqnarray}
where
\begin{eqnarray}
\Delta_{k}  =   \lambda \left< d_{k\uparrow}^{\dagger}d_{-k \downarrow}^{\dagger}\right>
\end{eqnarray}
is the superconducting order parameter of the layer. This Hamiltonian has also been recently studied by Mitra~\cite{mitra}.
\section{Keldysh formalism for the Green's functions} 

In this section we introduce the Keldysh formalism~\cite{keldysh, kadanoff} to obtain the normal and anomalous Green's functions relevant for the problem. The use of the Keldysh approach~\cite{keldysh, kadanoff}  is necessary due to the explicit time dependence of the kinetic term in the Hamiltonian $H_{layer}$ when the electric field is present. A direct consequence of this time dependence is that the Green's functions are now functions of $(t ,t^{\prime})$ and not of $(t -t^{\prime})$
 as in the case $\vec{A}=0$.

The normal and anomalous Green's functions are given respectively by (we follow the notation of Ref.~\cite{tyablikov}),
\begin{eqnarray}
\ll d_{k \sigma}(t)|d_{k \sigma}^{\dagger}(t^{'})\gg &\equiv &G^{c}(t,t^{'}) \\
\ll d_{-k -\sigma}^{\dagger}(t)|d_{k \sigma}^{\dagger}(t^{'})\gg &\equiv &G^{pc}(t,t^{'})
\end{eqnarray}
The retarded and advanced Green's functions obey the equations of motion,
\begin{eqnarray}
\left(i \hbar \frac{\partial}{\partial t} \!+\! \epsilon (t) \right) G^{pc}(t,t^{'}) &=& -  \Delta_{k}(t) G^{c}(t,t^{'}) - \sum_{k_{z}} t_z  \ll c_{k_{z} -k -\sigma}^{\dagger}(t)|d_{k\sigma}^{\dagger}(t^{'})\gg \nonumber\\
\left(i \hbar \frac{\partial}{\partial t}\! -\! \epsilon (t) \right) G^{c}(t,t^{'}) &=& \delta (t\!-\!t^{'}) \!-\!  \Delta_{k}(t) G^{pc}(t,t^{'})\! +\! \sum_{k_{z}} t_z^{*}  \ll \!c_{k_{z} k \sigma}(t)|d_{k\sigma}^{\dagger}(t^{'})\!\gg , \nonumber
\end{eqnarray}
where we defined 
$$
\epsilon \left( k-\frac{e\vec{A}(t)}{\hbar c}\right) \equiv \epsilon (t).
$$
The new  Green's functions  obey the equations of motion,
\begin{eqnarray}
\left(i \hbar \frac{\partial}{\partial t} + \epsilon _{k_{z},k,\sigma}^{b}\right) \ll c_{k_{z} -k -\sigma}^{\dagger}(t)|d_{k\sigma}^{\dagger}(t^{'})\gg &=& - t_z ^{*} G^{pc}(t,t^{'}) \nonumber\\
\left(i \hbar \frac{\partial}{\partial t} - \epsilon_{k_{z},k,\sigma}^{b} \right) \ll c_{k_{z} k \sigma}(t)|d_{k\sigma}^{\dagger}(t^{'})\gg &=&  t_z G^{c}(t,t^{'}). \nonumber
\end{eqnarray}
where $\epsilon _{k_{z},k,\sigma}^{b}$ is the kinetic energy of the electrons in the substrate (bath).
To solve the system above, we introduce four auxiliary Green's functions through their equations of motion.
\begin{eqnarray}
\label{auxiliar1}
\left(i \hbar \frac{\partial}{\partial t} + \epsilon _{k_{z},k,\sigma}^{b}\right) g_{k_{z} k \sigma}(t-t^{'})  &=& \delta (t-t^{'}) \\
\label{auxiliar2}
\left(i \hbar \frac{\partial}{\partial t} - \epsilon _{k_{z},k,\sigma}^{b}\right) g_{k_{z} k \sigma}^{'}(t-t^{'})  &=&  \delta (t-t^{'}) \\
\left(i \hbar \frac{\partial}{\partial t} + \epsilon (t) \right) g^{pc}(t,t^{'}) &=& \delta (t-t^{'}) \\
\left(i \hbar \frac{\partial}{\partial t} - \epsilon (t) \right) g^{c}(t,t^{'}) &=& \delta (t-t^{'}) , 
\end{eqnarray}
such that, 
\begin{eqnarray}
 \ll c_{k_{z} -k -\sigma}^{\dagger}(t)|d_{k\sigma}^{\dagger}(t^{'})\gg &=& -\int dt_{1} t_z^{*} g_{k_{z} k \sigma}(t, t_{1})G^{pc} (t_{1},t^{'}) \\
\ll c_{k_{z} k \sigma}(t)|d_{k\sigma}^{\dagger}(t^{'})\gg &=& \int dt_{1} t_z g_{k_{z} k \sigma}^{'}(t, t_{1})G^{c} (t_{1},t^{'}) 
\end{eqnarray}
and
\begin{eqnarray}
 G^{pc} (t,t^{'})&=& -  \int \! dt_{1} g^{pc}(t, t_{1})\Delta_{k}(t_1)G^{c} (t_{1},t^{'}) -\nonumber\\
&&- \sum_{k_{z}}\! t_z\! \int\! dt_{1} g^{pc}(t, t_{1})\ll c_{k_{z} -k -\sigma}^{\dagger}(t_{1})|d_{k\sigma}^{\dagger}(t^{'})\gg \\
G^{c} (t,t^{'}) &=& g^{c}(t,t^{'}) -  \int dt_{1} t_z^{*} g^{c}(t, t_{1})\Delta_{k}(t_1)G^{c} (t_{1},t^{'}) + \nonumber\\
&&+\sum_{k_{z}} t_z^{*} \int dt_{1} g^{c}(t, t_{1})\ll c_{k_{z} k \sigma}(t_{1})|d_{k\sigma}^{\dagger}(t^{'})\gg. 
\end{eqnarray}
Next, we define the self-energies as:
\begin{eqnarray}
\Sigma(t_1,t_2) = \sum_{k_{z}} |t_z|^{2} g_{k_{z} k \sigma}(t_1,t_2) \approx \bar{\Sigma} (t_1,t_2)= \sum_{k_{z}} |t_z|^{2} g_{k_{z} k \sigma}^{'}(t_1,t_2). \label{self}
\end{eqnarray}
In terms of these self-energies, we have the following implicit equations for the anomalous and normal Green's functions,
\begin{eqnarray}
\label{final1}
G^{pc}(t, t^{'}) &=& -\!  \int dt_{1} g^{pc}(t, t_{1})\Delta _{k}(t_1)G^{c}(t_{1}, t^{'}) + \nonumber\\
&& + \int dt_{1} \int dt_{2} g^{pc}(t,t_{1}) \Sigma (t_{1}, t_{2}) G^{pc}(t_{2}, t^{'}) 
\end{eqnarray}
\begin{eqnarray}
\label{final2}
G^{c}(t, t^{'}) &=& g^{c}(t, t^{'}) \!-\! \int dt_{1} g^{c}(t, t_{1})\Delta _{k}(t_1) G^{pc}(t_{1}, t^{'}) + \nonumber\\
&& + \int dt_{1} \int dt_{2} g^{c}(t,t_{1}) \Sigma (t_{1}, t_{2}) G^{c}(t_{2}, t^{'}). 
\end{eqnarray}

\section{Adiabatic approximation}

To make further progress we make a reparametrization in time of the equations above. A convenient choice is to introduce a \textit{slow time}  [$(t+t^{'})/2$] and a \textit{fast time} ($t-t^{'}$)~\cite{alexis}, such that,
\begin{eqnarray}
(t, t^{'}) \rightarrow \left( t-t^{'}, \frac{t+t^{'}}{2}\right) = (t-t^{'}, \bar{t}). \nonumber
\end{eqnarray}
In terms of these \textit{times}, equations (\ref{final1}) and (\ref{final2}) become,
\begin{eqnarray}
\label{Gpc}
&&G^{pc}(t-t^{'}, \bar{t}) = -  \int dt_{1} g^{pc}\left( t-t_{1}, \frac{t+t_{1}}{2}\right)\Delta _{k}(t_{1})G^{c}\left(t_{1}-t^{'}, \frac{t_{1}+t^{'}}{2}\right) \nonumber\\
&&+ \int dt_{1} \int dt_{2} g^{pc}\left( t-t_{1}, \frac{t+t_{1}}{2}\right) \Sigma \left(t_{1}-t_{2}, \frac{t_{1}+t_{2}}{2}\right) G^{pc}\left(t_{2}-t^{'}, \frac{t_{2}+t^{'}}{2}\right) \nonumber\\
\end{eqnarray}
\begin{eqnarray}
\label{Gc}
&&G^{c}(t-t^{'}, \bar{t}) = g^{c}(t-t^{'}, \bar{t}) \!-\!  \int dt_{1} g^{c}\left( t\!-\!t_{1}, \frac{t\!+\!t_{1}}{2}\right)\Delta_k(t_{1})G^{pc}\left(t_{1}\!-\!t^{'}, \frac{t_{1}\!+\!t^{'}}{2}\right) \nonumber\\
&&+ \int dt_{1} \int dt_{2} g^{c}\left( t-t_{1}, \frac{t+t_{1}}{2}\right) \Sigma \left(t_{1}-t_{2}, \frac{t_{1}+t_{2}}{2}\right) G^{c}\left(t_{2}-t^{'}, \frac{t_{2}+t^{'}}{2}\right). \nonumber\\
\end{eqnarray}
Once we have distinguished between the fast and slow time scales we can implement an  adiabatic approximation~\cite{alexis,dissertacao} which in our case consists to take into account terms up to linear order in the variables associated with the slow time scale. The validity of this approximation requires the characteristic time associated with the change of the external parameter (the vector potential)  to be large compared to the lifetime of an electron in the superconducting layer before it is scattered to the metallic substrate.  Since the time derivative of the vector potential is related to the electric field, the adiabatic condition turns out to involve the electric field. This condition will be given and discussed in more detail below.

Expanding  the relevant Green's functions to linear order in the variables associated with the slow time scale, 
{\small{
\begin{eqnarray}
\label{exp}
G\left(t-t^{'}, \frac{t\!+\!t^{'}}{2}\right) \approx G(t-t^{'}, \bar{t}) \!+\! \left(\frac{t^{'} - t}{2}\right) \frac{\partial}{\partial \bar{t}} \left.G(t\!-\!t^{'}, \bar{t}) \right|_{\bar{t}=t} + O[(\bar{t}\!-\!t)^{2}],
\end{eqnarray}
}}
which is rewritten as,
\begin{eqnarray}
\label{order}
G(t-t^{\prime}, \bar{t})=G^{(0)}(t-t^{\prime}, \bar{t})+G^{(1)}(t-t^{\prime}, \bar{t}),
\end{eqnarray}
where the zeroth order term refers to equilibrium quantities.

Using the expansion (\ref{exp}) in equations (\ref{Gpc}) and (\ref{Gc}), up to linear order in the slow variable, we get, 
\begin{eqnarray}
G^{pc}(t - t^{'}, \bar{t}) &=& -  \Delta_{k}(\bar{t}) \int  dt_{1}  \left[ g^{pc}(t - t_{1}, \bar{t})G^{c}(t_{1} - t^{'}, \bar{t}) +\right. \nonumber\\ 
&& +   \left(\frac{t_{1} - t}{2}\right) g^{pc}(t - t_{1}, \bar{t}) \frac{\partial G^{c}(t_{1}-t^{'}, \bar{t})}{\partial \bar{t}} + \nonumber\\
&&+ \left. \left( \frac{t_{1}-t^{'}}{2} \right) \frac{\partial g^{pc}(t-t_{1}, \bar{t})}{\partial \bar{t}} G^{c}(t_{1}-t^{'}, \bar{t}) \right] -\nonumber\\
&& -   \int dt_{1}  g^{pc}(t-t_{1}, \bar{t})(t_1-\bar{t}) \frac{\partial \Delta_k(\bar{t})}{\partial \bar{t}} G^{c}(t_{1}-t^{'}, \bar{t})  +\nonumber\\
&&+  \int dt_{1} \int  dt_{2} \left[ g^{pc}(t - t_{1}, \bar{t})\Sigma(t_{1} - t_{2}, \bar{t})G^{pc}(t_{2} - t^{'}, \bar{t})  + \right. \nonumber\\
&& +  \left(\frac{t_{2}- t}{2} \right) g^{pc}(t-t_{1}, \bar{t}) \Sigma(t_{1} - t_{2}, \bar{t}) \frac{\partial G^{pc}(t_{2}-t^{'}, \bar{t})}{\partial \bar{t}} + \nonumber\\
&&+  \left(\frac{t_{1}-t}{2}+ \frac{t_{2}-t^{'}}{2}\right) g^{pc}(t-t_{1}, \bar{t}) \frac{\partial}{\partial \bar{t}}  \left(\Sigma (t_{1}-t_{2}, \bar{t}) G^{pc}(t_{2}-t^{'}, \bar{t})\right) + \nonumber\\
&& \left. + \left(\frac{t_{1} - t^{'}}{2}\right) \frac{\partial}{\partial \bar{t}}\left[g^{pc}(t - t_{1}, \bar{t}) \Sigma (t_{1} - t_{2}, \bar{t})\right] G^{pc}(t_{2} - t^{'}, \bar{t})\right] 
\end{eqnarray}
and
\begin{eqnarray}
G^{c}(t - t^{'}, \bar{t}) &=& g^{c}(t-t^{'}, \bar{t}) -  \Delta_{k}(\bar{t}) \int dt_{1} \left[g^{c}(t-t_{1}, \bar{t})G^{pc}(t_{1}-t^{'}, \bar{t}) +\right. \nonumber\\
&& +\left(\frac{t_{1}- t}{2}\right) g^{c}(t- t_{1}, \bar{t}) \frac{\partial G^{pc}(t_{1} - t^{'}, \bar{t})}{\partial \bar{t}}  +\nonumber\\
&& + \left.  \left(\frac{t_{1}-t^{'}}{2}\right) \frac{\partial g^{c}(t-t_{1}, \bar{t})}{\partial \bar{t}} G^{pc}(t_{1}-t^{'}, \bar{t}) \right] - \nonumber\\
&& -    \int dt_{1} g^{c}(t-t_{1}, \bar{t})(t_{1}-\bar{t})\frac{\partial \Delta_{k}(\bar{t})}{\partial \bar{t}}G^{pc}(t_{1}-t^{'}, \bar{t}) + \nonumber\\
&& + \int dt_{1} \int dt_{2} \left[ g^{c}(t-t_{1}, \bar{t})\Sigma(t_{1}-t_{2}, \bar{t})G^{c}(t_{2}-t^{'}, \bar{t}) +\right. \nonumber\\
&&  + \left(\frac{t_{2}-t}{2}\right)g^{c}(t-t_{1}, \bar{t}) \Sigma(t_{1}-t_{2}, \bar{t}) \frac{\partial G^{c}(t_{2}-t^{'}, \bar{t})}{\partial \bar{t}} + \nonumber\\
&& \left.+ \left( \frac{t_{1}-t}{2}+\frac{t_{2}-t^{'}}{2} \right) g^{c} (t-t_{1}, \bar{t}) \frac{\partial}{\partial \bar{t}} \left( \Sigma ( t_{1}-t_{2}, \bar{t} ) G^{c}( t_{2} - t^{'}, \bar{t} ) \right)+  \right. \nonumber\\
&&+ \left. \left(\frac{t_{1}-t^{'}}{2}\right) \frac{\partial}{\partial \bar{t}}\left(g^{c}(t-t_{1}, \bar{t}) \Sigma (t_{1}-t_{2}, \bar{t})\right) G^{c}(t_{2}-t^{'}, \bar{t})\right] 
\end{eqnarray}
where terms of second order in $(\bar{t} -t^{'})$ have been neglected.

Fourier transforming in the fast variables $(t- t^{\prime})$ and using the expansion Eq. (\ref{exp})  allows to separate the Green's functions and the order parameter in two contributions, respectively, in zero order and first order in the slow variable $\bar{t}$, as in Eq.~(\ref{order}). We can collect terms of the same order and finally obtain,
\begin{eqnarray}
&&G^{pc(0)} = - \Delta_{k}^{(0)}g^{pc(0)}G^{c(0)} \!+\! g^{pc(0)} \Sigma G^{pc(0)} \\
\nonumber\\
G^{pc(1)} &=& -  \Delta_{k}^{(0)} \left( g^{pc}G^{c}\right)^{1} -  \Delta_{k}^{(1)} g^{pc(0)} G^{c(0)} + \frac{i }{2} \frac{\partial}{\partial \bar{t}} \Delta_{k}^{(0)} \left( g^{pc(0)} G^{c(0)}\right)_{\partial_{\omega}}+ \nonumber\\
&& + \Sigma \left( g^{pc} G^{pc}\right)^{1} + \frac{i }{2} \Delta_{k}^{(0)} \left( g^{pc(0)} G^{c(0)}\right)_{\partial_{\omega} \partial_{\bar{t}}} - \frac{i}{2} \frac{\partial}{\partial \omega} \left( g^{pc(0)} \Sigma\right) \frac{\partial}{\partial \bar{t}} G^{pc(0)}+ \nonumber\\
&& + \frac{i}{2} \frac{\partial}{\partial \bar{t}} g^{pc(0)} \frac{\partial}{\partial \omega} \left( \Sigma G^{pc(0)}\right)
\end{eqnarray}
and
\begin{eqnarray}
&&G^{c(0)}= g^{c(0)} -  \Delta_{k}^{(0)}g^{c(0)}G^{pc(0)} + g^{c(0)} \Sigma G^{c(0)} \\
\nonumber\\
G^{c(1)}&=& g^{c(1)} \!-\!  \Delta_{k}^{(0)} \left( g^{c} G^{pc}\right)^{1} \!-\!  \Delta_{k}^{(1)} g^{c(0)} G^{pc (0)}  \!+\! \frac{i }{2} \frac{\partial}{\partial \bar{t}}\Delta_{k}^{(0)} \left( g^{c(0)} G^{pc (0)} \right)_{\partial_{\omega}} + \nonumber\\
&& + \Sigma \left( g^{c} G^{c}\right)^{1} + \frac{i}{2} \Delta_{k}^{(0)} \left( g^{c(0)} G^{pc (0)}\right)_{\partial_{\omega} \partial_{\bar{t}}} - \frac{i}{2} \frac{\partial}{\partial \omega} \left(g^{c(0)} \Sigma \right) \frac{\partial}{\partial \bar{t}}G^{c(0)} + \nonumber\\
&&+ \frac{i}{2} \frac{\partial}{\partial \bar{t}} g^{c(0)} \frac{\partial}{\partial\omega}(\Sigma G^{c(0)})
\end{eqnarray}
where the functions above depend on  $(\omega, \bar{t})$ and we have neglected terms of second order in $\bar{t} - t^{'}$. We also used that $\Sigma$ is time independent and introduced the following notation,
\begin{eqnarray}
&& (fg)^{1} = f^{0} g^{1} + f^{1} g^{0} \nonumber\\
&& (fg)_{\partial_{\omega} \partial_{\bar{t}}} = \partial_{\omega} f \partial_{\bar{t}} g - \partial_{\bar{t}} f \partial_{\omega} g \nonumber\\
&& (fg)_{\partial_{\omega (\bar{t})}} = g \partial_{\omega (\bar{t})} f - f \partial_{\omega (\bar{t})} g ,\nonumber
\end{eqnarray}
where $f$ and $g$ are any two Green's function.

We will now solve the problem above considering  terms of the same order separately. 

\subsection{Zero order solution} 

Let us consider initially the zero order problem. At this level of approximation the slow time is taken as fixed at an arbitrary time $\bar{t}=\bar{t}_0$, which for simplicity we take as zero ($\bar{t}_0=0$). This corresponds to take the electric field as zero, such that, there is no explicit time dependence in the problem. We want to obtain the  retarded, advanced and Keldysh (lesser) components of the Green's functions. For the retarded and advanced Green's functions in zero order, we get,

\begin{eqnarray}
 \left\{ 
\begin{array}{ll}
G^{pc(0)r(a)} \!=\! - \Delta_{k}^{(0)}g^{pc(0)r(a)}G^{c(0)r(a)} \!+\! g^{pc(0)r(a)} \Sigma^{r(a)} G^{pc(0)r(a)} \\
G^{c(0)r(a)}= g^{c(0)r(a)} -  \Delta_{k}^{(0)}g^{c(0)r(a)}G^{pc(0)r(a)} + g^{c(0)r(a)} \Sigma^{r(a)} G^{c(0)r(a)} ,
\end{array} \right.
\end{eqnarray}
which are functions of  $(\omega, \bar{t})$. Notice that we have a system of two equations and two unknowns that can be easily solved. 

Once we find $G^{c(0)r(a)}$ e $G^{pc(0)r(a)}$, we can use the fluctuation-dissipation theorem for the zero order Green's functions to obtain the Keldysh (lesser)  anomalous Green's function. We get,
\begin{eqnarray}
G^{pc(0)<} = f(\omega) [G^{pc(0)a} -G^{pc(0)r}].
\end{eqnarray}

Since $g^{c(0)r(a)}$ and $g^{pc(0)r(a)}$ given by Eqs.~(\ref{auxiliar1}) and (\ref{auxiliar2}) can be directly obtained, we can use again the fluctuation-dissipation theorem to obtain a complete solution of the system of equations. For the self-energies we use the following approximation~\cite{mitra},
\begin{eqnarray}
\label{autoe}
\Sigma^{r(a)}  \approx  \bar{\Sigma } \approx \mp i \Gamma, \nonumber\\
\end{eqnarray}
where the signs $(-)$ and $(+)$ refer to the retarded and advanced components of the self-energy, respectively. The physical meaning of $\Gamma$ is the inverse of the lifetime $\tau=1/\Gamma$ that the electron stays in the superconducting film before being scattered to the substrate.
Using the fluctuation-dissipation theorem to obtain $\Sigma^{<}$, we finally get,
\begin{eqnarray}
G^{pc(0)r(a)} = \frac{ \Delta_{k}^{(0)}}{(\Gamma \mp i \omega)^{2} + E_{k}^{2}}
\end{eqnarray}
and 
\begin{eqnarray}
\label{menorordem0}
G^{pc(0)<} = - \frac{4 i \Gamma  \Delta_{k}^{(0)}\omega f(\omega)}{\left[ \Gamma^{2} - \omega^{2} + E_{k}^{2}\right]^{2} + 4 \Gamma^{2} \omega^{2}},
\end{eqnarray}
where $E_{k} = \sqrt{\Delta_{k}^{(0)2} + \epsilon^{2}_{k}}$. Similarly we can obtain solutions for the Green's functions $G^{c(0)r(a)}$ e $G^{c(0)<}$.

From Eq.~(\ref{menorordem0}) and using that,
\begin{eqnarray}
\Delta_{k}^{(0)} = \lambda \sum_{k} \int \frac{1}{2 \pi i} G^{pc(0)<} (\omega, \bar{t}) d \omega,
\label{int1}
\end{eqnarray}
we can find the gap equation for $\Delta_{k}^{(0)}$. This is given by, 
\begin{eqnarray}
\label{intd0}
\Delta_{k}^{(0)}  = \sum _{k} \frac{\lambda}{\pi} \int d\omega f(\omega) \frac{ \Delta_{k}^{(0)}}{E_{k}} \left[- \frac{\Gamma }{(\omega - E_{k})^{2} +  \Gamma ^{2} } + \frac{\Gamma }{(\omega + E_{k})^{2} + \Gamma ^{2} }\right]. 
\end{eqnarray}
This equation can be solved self-consistently to yield the superconducting order parameter as a function of temperature and of the coupling to the metallic substrate through the damping parameter $\Gamma$. Since here we are interested in obtaining the finite temperature phase diagram of the system, we take $\Delta_{k}^{(0)}=0$.  Performing the integration in $\omega$, we get:
\begin{equation}
\label{temp}
\frac{1}{\lambda \nu } =  \frac{1}{  \pi} \int_{- \Omega}^{\Omega} \frac{d \epsilon}{ \epsilon} Im \left[ \psi \left( \frac{1}{2} + \frac{ \beta_c (\Gamma_c + i \epsilon)}{2 \pi} \right) \right]
\end{equation}
where $\beta_c=1/k_B T_c$, $\nu$ is the density of states at the Fermi level and $\psi$ the digamma function. It can be easily checked that when $\Gamma = 0$, this yields the usual mean-field BCS result for the critical temperature $T_c$.
On the other hand, at zero temperature we find there is a critical value of the dissipation parameter $\Gamma_c$ above which superconductivity in the layer is destroyed. This is given by,
\begin{eqnarray}
\label{eqmitra}
\frac{1}{\lambda \nu }  =   \ln \left(\frac{\Omega}{\Gamma_{c}}\right) \Longrightarrow \Gamma_{c} = \Omega \exp\left[-\frac{1}{ \nu}\right] , 
\end{eqnarray}
or
\begin{eqnarray}
\label{gamac}
\Gamma_{c} = \frac{\Delta_{0}}{2},
\end{eqnarray}
where $\Delta_{0} = 2 \Omega e^{-1/ \lambda \nu}$ is the gap in the 2D layer in equilibrium.

The critical value of the dissipation, Eq.~(\ref{gamac}), and the zero order  phase diagram of the superconducting film with dissipation $T_c(\Gamma)$ coincide with the results obtained by Mitra~\cite{mitra} using a different approach.

\subsubsection{Nature of the dissipation induced transition ($\vec{A}=0$)}

We have shown above that for a sufficiently strong coupling between the layer and the normal substrate, superconductivity can be suppressed at a quantum critical point (QCP) at $\Gamma_c$. An important question, for which the mean-field approach above can not give an appropriate answer concerns the nature of this transition. Which is the universality class of the dissipation induced QCP and in particular which is the value of its associated dynamic exponent $z$?
To investigate this problem we need to include fluctuations to the zero order solution. For this purpose we use an alternative approach which consists to calculate the response of the layer, in the absence of the vector potential, to a fictitious, frequency and wavevector dependent field of intensity $h$ that couples to the superconducting order parameter~\cite{aline}. This response is given in terms of a generalized susceptibility $\chi(q,\omega)$, such that,
\begin{eqnarray}
\delta \Delta^{(0)}_k (\omega) = \frac{\chi (q, \omega)}{1-  \lambda \chi(q, \omega)}h.
\end{eqnarray}
The condition
\begin{eqnarray}
1-  \lambda \chi(q=0, \omega=0) = 0.
\end{eqnarray}
signals an instability to a homogeneous superconducting state, since $\Delta^{(0)}$ can be finite even in the absence of the fictitious field. On the other hand,  considering the frequency and wave vector dependence of the generalized susceptibility amounts to take into account the fluctuations close to this instability.
The quantity $\chi(q, \omega)$ is the q-dependent dynamic pair susceptibility~\cite{Thouless} of the layer, that we generalize here to include a finite life-time of the quasi-particles in the normal phase. It is given by,
\begin{eqnarray}
\label{susceptibilidade}
\chi (q, \omega) = \frac{1}{2} \sum_{k} \left[ \frac{\tanh(\beta \epsilon_{k}/2)}{\epsilon_{k+q}+\epsilon_{k} - (\omega + i \tau_{SC}^{-1}/2)} + \frac{\tanh(\beta \epsilon_{k}/2)}{\epsilon_{k-q}+\epsilon_{k} - (\omega + i \tau_{SC}^{-1}/2)} \right],
\end{eqnarray}
where $\beta = (k_{B}T)^{-1}$. Dissipation is included in the term $i \tau_{SC}^{-1}$. We take $\tau_{SC}=\tau=\Gamma^{-1}$ which is the life-time of the quasi-particles in the layer due to the coupling with the metallic substrate.
For $T=0$ and close to $q=0$ and $\omega=0$, we can calculate this susceptibility and obtain,
\begin{eqnarray}
\chi(q\approx 0, \omega \approx 0) = \frac{\nu}{4} \ln \left(\frac{\Omega^{4}}{\Gamma^{4}}\right) - \frac{\nu (q v_{F})^{2}}{2 \Gamma^{2}} + i\frac{\nu \omega}{\Gamma} .
\end{eqnarray}
Notice that the relevant wave-vector to expand is near $q=0$ since we are interested in an instability to a uniform superconducting state. The Thouless condition $1- \Re e \chi(q=0,\omega=0) = 0$ yields,
\begin{eqnarray}
\frac{1}{ \lambda \nu} = \ln \left( \frac{\Omega}{\Gamma_{c}} \right).
\end{eqnarray}
which coincides with Eq.~(\ref{eqmitra}) obtained previously. Consequently, the present approach that incorporates the finite life-time of the quasi-particles in the layer through the dynamic pair susceptibility yields the same dissipation induced quantum critical point obtained previously. The advantage of course is that we have now a full dynamic description of the quantum phase transition. Following the approach of Hertz~\cite{hertz}, we can write an effective action, at the Gaussian level, which describes this QCP,
\begin{eqnarray}
S = \int dq \int d\omega \left[  \left( \frac{\Gamma - \Gamma_{c}}{\Gamma_{c}} \right) + \frac{ (q v_{F})^{2}}{2 \Gamma^{2}} + \frac{ |\omega |}{\Gamma}\right]|\Delta^{0}(q, \omega)|^2.
\end{eqnarray}
where $\Gamma_c=\Delta_0/2$ as before. The dynamic exponent turns out to be $z = 2$ and the effective dimension of the QCP,  $d_{eff} = d + z=4$~\cite{livro}. Then the quantum  normal-to-superconductor phase transition in the layer when its coupling to the metallic substrate is reduced occurs at the upper critical dimension $d_c=4$ in which case logarithmic corrections to the Gaussian or mean-field critical behavior is expected~\cite{livro}.

Later on we will use the the properties of this QCP, in particular the values of the critical exponents $\nu=1/2$ and $z=2$, to discuss the phase diagram and scaling properties in the presence of interactions and an electric field.

\subsection{First order approach}

Let us now consider the first order terms of the Green's functions and of the order parameter.
First, we notice that due to the time reparametrization, we can easily show that the Green's functions
 $g^{pc(1) r(a)<} = g^{c(1) r(a)<} = 0$  and  $g^{pc(1)} = g^{c(1)} = 0$~\cite{dissertacao}.  Using this simplification in equations (30) and (32) and applying the Langreth rules~\cite{langreth, alexis} for  the retarded and advanced components, we get a system with two equations and two parameters. We can solve this system and find:
\begin{eqnarray}
\label{c1}
G^{c (1) r(a)} &=& \frac{1-g^{pc(0) r(a)}}{\left( 1 - g^{c(0) r(a)}\Sigma ^{r(a)}\right)\left( 1- g^{pc(0)r(a)}\Sigma ^{r(a)}\right) -  ^{2}\Delta_{k}^{(0) 2}g^{pc(0)r(a)}g^{c(0) r(a)}}\times \nonumber\\
&& \left\{ -  \Delta_{k}^{(0)} \left[ \frac{g^{c(0)r(a)}}{1 - g^{pc(0)r(a)}\Sigma ^{r(a)}} \right. \right. \left\{ \frac{ \Delta_{k}^{(0)}}{2} \left( g^{pc(0) r(a)} G^{c(0) r(a)}\right)_{\partial_{\omega} \partial_{\bar{t}}} \right. - \nonumber\\
&& -  \Delta_{k}^{(1)} g^{pc(0)r(a)} G^{c(0)r(a)} - \frac{1}{2} \frac{\partial}{\partial \omega} \left( g^{pc(0)r(a)} \Sigma^{r(a)}\right)\frac{\partial}{\partial \bar{t}} G^{pc(0)r(a)} + \nonumber\\
&& \left.+ \frac{i}{2} \frac{\partial}{\partial \bar{t}} g^{pc(0)r(a)} \frac{\partial}{\partial \omega} \left( \Sigma^{r(a)} G^{pc(0)r(a)}\right) \right\} - \frac{i }{2} \left(g^{pc(0)r(a)} G^{c(0)r(a)} \right)_{\partial_{\omega}} - \nonumber\\
&&  \left.  - \frac{i}{2} \left( g^{c(0)r(a)} G^{pc(0)r(a)}\right)_{\partial_{\omega} \partial_{\bar{t}}} \right] -  \Delta_{k}^{(1)}g^{c(0) r(a)} G^{pc(0)r(a)} +\nonumber\\
&&  + \! \frac{ i}{2} \frac{\partial \Delta_{k}^{(0)}}{\partial \bar{t}} \left( g^{c(0)r(a)} G^{pc(0)r(a)} \right)\!_{\partial \omega} \!-\! \frac{i}{2}\frac{\partial}{\partial \omega}\left( g^{c(0)r(a)} \Sigma^{r (a)} \right) \frac{\partial}{\partial \bar{t}} G^{c(0)r(a)} \! + \nonumber\\
&& \left. + \frac{1}{2} \frac{\partial}{\partial \bar{t}} g^{c(0)r(a)}\frac{\partial}{\partial \omega}\left( \Sigma^{r(a)} G^{c(0)r(a)}\right) \right\}
\end{eqnarray}
that depends only on known quantities.

Also,
\begin{eqnarray}
\label{pc1}
G^{pc(1)r(a)} &=& \frac{1}{1\!-\!\Sigma^{r(a)} g^{pc(0)r(a)}} \left\{-\!  \Delta_{k}^{(0)} g^{pc(0) r(a)} G^{c (1) r(a)}\!-\!\frac{i}{2}\left( g^{pc (0)r(a)} G^{c (0)r(a)} \right)_{\partial_{\omega} \partial_{\bar{t}}} \!- \right.\nonumber\\
&& -  \Delta_{k}^{(1)} g^{pc(0)r(a)} G^{c(0)r(a)} - \frac{i}{2}\frac{\partial}{\partial \omega}\left( g^{pc(0)r(a)}\Sigma^{r(a)}\right)\frac{\partial}{\partial \bar{t}} G^{pc(0) r(a)}+ \nonumber\\
&& \left. +\! \frac{i}{2}\frac{\partial}{\partial \bar{t}} g^{pc(0)r(a)} \frac{\partial}{\partial \omega}\left( g^{pc(0)r(a)}\Sigma^{r(a)}\right) \!+\! \frac{i }{2} \frac{\partial \Delta_{k}^{(0)}}{\partial \bar{t}} \left( g^{pc(0)r(a)} G^{c(0)r(a)}\right)_{\partial_{\omega}} \right\} \nonumber\\
\end{eqnarray}
that involves $G^{pc(1)r(a)}$ calculated before (Eq.~(\ref{pc1})). In equations (\ref{c1})  and (\ref{pc1}), we use the fact that the self-energy is independent of  $\bar{t}$ (see Eq.~\ref{autoe}).
Using the Langreth rules~\cite{langreth, alexis} for the retarded and advanced  components of the Green's functions, we get for the first order Keldysh (lesser) Green's function:
\begin{eqnarray}
G^{c(1)<} &=&  \frac{\left(1-g^{pc(0)r} \Sigma^{r}\right)}{\left( 1 - g^{c(0) r}\Sigma ^{r}\right)\left( 1- g^{pc(0)r}\Sigma ^{r}\right) -  ^{2}\Delta_{k}^{(0) 2}g^{pc(0)r}g^{c(0) r}} \left\{  \frac{- \Delta_{k}^{(0)} g^{c(0)r}}{1-g^{pc(0)r}\Sigma^{r}}  \right.  \nonumber\\
&&  \left\{ \Delta_{k}^{(0)} \frac{i}{2} \left( g^{pc (0) r} G^{c (0) < }\right)_{\partial_{\omega} \partial_{\bar{t}}} +  \Delta_{k}^{(0)} \frac{i}{2} \left( g^{pc (0) <} G^{c (0) a }\right)_{\partial_{\omega} \partial_{\bar{t}}} - \right. \nonumber\\
&& - \Delta_{k}^{(0)}g^{pc(0)<}G^{c(1)a} -   \Delta_{k}^{(1)}g^{pc(0)r}G^{c(0)<} - \!  \Delta_{k}^{(1)}g^{pc(0)<}G^{c(0)a} \!+ \nonumber\\
&&   +\! g^{pc(0)r}\Sigma^{<}G^{pc(1)a} \!+\! g^{pc(0)<}\Sigma^{a}G^{pc(1)a} \!+ \frac{i }{2} \frac{\partial \Delta_{k}^{(0)}}{\partial \bar{t}} \left( g^{pc (0) r} G^{c(0)<}\right)_{\partial \omega} + \nonumber\\
&& + \frac{i }{2} \frac{\partial \Delta_{k}^{(0)}}{\partial \bar{t}} \left( g^{pc (0) <} G^{c(0)a}\right)_{\partial \omega} - \frac{i}{2}\frac{\partial}{\partial \omega}\left( g^{pc(0)r}\Sigma^{r}\right)\frac{\partial G^{pc(0)<}}{\partial \bar{t}}  - \nonumber\\
&& - \frac{i}{2}\frac{\partial}{\partial \omega}\left( g^{pc(0)r}\Sigma^{<}\right)\frac{\partial G^{pc(0)a}}{\partial \bar{t}} - \frac{i}{2}\frac{\partial}{\partial \omega}\left( g^{pc(0)<}\Sigma^{a}\right)\frac{\partial G^{pc(0)a}}{\partial \bar{t}} + \nonumber\\
&& + \frac{i}{2} \frac{\partial g^{pc(0)r}}{\partial \bar{t}}\frac{\partial}{\partial \omega}\left( \Sigma^{r} G^{pc(0)<}\right) + \frac{i}{2} \frac{\partial g^{pc(0)r}}{\partial \bar{t}}\frac{\partial}{\partial \omega} \left( \Sigma^{<} G^{pc(0)a}\right) + \nonumber\\
&& \left. + \frac{i}{2} \frac{\partial g^{pc(0)<}}{\partial \bar{t}}\frac{\partial}{\partial \omega} \left( \Sigma^{a} G^{pc(0)a}\right) \right\} -  \Delta_{k}^{(0)} g^{c(0)<}G^{pc(1)a} -  \Delta_{k}^{(1)}g^{c(0)<}G^{pc(0)a} + \nonumber\\
&&+ \frac{i  \Delta_{k}^{(0)}}{2} \left( g^{c (0) r} G^{pc (0)<}\right)_{\partial_{\omega} \partial_{\bar{t}}} + \frac{i  \Delta_{k}^{(0)}}{2} \left( g^{c (0) <} G^{pc (0)a}\right)_{\partial_{\omega} \partial_{\bar{t}}} -\nonumber\\
&&   -  \Delta_{k}^{(1)}g^{c(0)r}G^{pc(0)<} + g^{c(0)r}\Sigma^{<}G^{c(1)a} + g^{c(0)<}\Sigma^{a}G^{c(1)a}  + \nonumber\\
&& + \frac{i }{2} \frac{\partial \Delta_{k}^{(0)}}{\partial \bar{t}} \left( g^{c (0) r} G^{pc (0) <}\right)_{\partial \omega} + \frac{i }{2} \frac{\partial \Delta_{k}^{(0)}}{\partial \bar{t}} \left( g^{c (0) <} G^{pc (0) a}\right)_{\partial \omega} - \nonumber\\
&&- \frac{i}{2} \frac{\partial g^{c(0)<}}{\partial \bar{t}} \frac{\partial}{\partial \omega} \left( \Sigma^{a} G^{c(0)a} \right) -  \frac{i}{2}\frac{\partial}{\partial \omega}\left(g^{c(0)r} \Sigma^{r} \right) \frac{\partial G^{c(0)<}}{\partial \bar{t}} + \nonumber\\
&&  + \frac{i }{2} \frac{\partial \Delta_{k}^{(0)}}{\partial \bar{t}} \left(g^{c (0) r} G^{pc (0)<} \right) + \frac{i }{2} \frac{\partial \Delta_{k}^{(0)}}{\partial \bar{t}} \left(g^{c (0) <} G^{pc (0)a} \right)  - \nonumber\\
&& - \frac{1}{2}\frac{\partial}{\partial \omega}\left(g^{c(0)r} \Sigma^{<} \right) \frac{\partial G^{c(0)a}}{\partial \bar{t}} - \frac{1}{2}\frac{\partial}{\partial \omega}\left(g^{c(0)<} \Sigma^{a} \right) \frac{\partial G^{c(0)a}}{\partial \bar{t}} + \nonumber\\
&& +\left.\frac{i}{2} \frac{\partial g^{c(0)r}}{\partial \bar{t}} \frac{\partial}{\partial \omega} \left( \Sigma^{r} G^{c(0)<} \right) + \frac{i}{2} \frac{\partial g^{c(0)r}}{\partial \bar{t}} \frac{\partial}{\partial \omega} \left( \Sigma^{<} G^{c(0)a} \right) \right\}
\end{eqnarray}
and
\begin{eqnarray}
\label{impo}
G^{pc(1)<} &=&  \frac{1}{1\!-\! \Sigma^{r} g^{pc (0) r}} \left[ \!-\! \Delta_{k}^{(0)} g^{pc (0)r} G^{c (1) <} \!-\!  \Delta_{k}^{(0)} g^{pc (0) <} G^{c (1) a} \!+\! g^{pc (0) <} \Sigma^{r} G^{pc(1)a}\right. + \nonumber\\
&&+\! g^{pc (0) a} \Sigma^{<} G^{pc (1)a} \!+\! \frac{i }{2} \frac{\partial \Delta_{k}}{\partial \bar{t}} \left( g^{pc (0) r} G^{c (0) <}\right)_{\partial_{\omega}} \!+\!  \frac{i }{2} \frac{\partial \Delta_{k}}{\partial \bar{t}} \left( g^{pc (0) <} G^{c (0) a}\right)\!_{\partial_{\omega}} \!+ \nonumber\\
&& + \frac{i}{2} \Delta_{k}^{(0)} \left( g^{pc (0) r} G^{c (0) <}\right)_{\partial_{\omega} \partial_{\bar{t}}} + \frac{i}{2} \Delta_{k}^{(0)} \left( g^{pc (0) <} G^{c (0) a}\right)_{\partial_{\omega} \partial_{\bar{t}}} - \nonumber\\
&& - \frac{1}{2} \frac{\partial}{\partial \omega} \left( g^{pc(0) r} \Sigma^{r}\right) \frac{\partial}{\partial \bar{t}} G^{pc (0) <} - \frac{1}{2} \frac{\partial}{\partial \omega} \left( g^{pc(0) r} \Sigma^{<}\right) \frac{\partial}{\partial \bar{t}} G^{pc (0) a} - \nonumber\\
&&- \frac{1}{2} \frac{\partial}{\partial \omega} \left( g^{pc(0) r<} \Sigma^{a}\right) \frac{\partial}{\partial \bar{t}} G^{pc (0) a} + \frac{i}{2} \frac{\partial}{\partial \bar{t}} g^{pc (0) r} \frac{\partial}{\partial \omega}\left(\Sigma^{r} G^{pc(0)<} \right) + \nonumber\\
&&+\left. \frac{i}{2} \frac{\partial}{\partial \bar{t}} g^{pc (0) r} \frac{\partial}{\partial \omega}\left(\Sigma^{<} G^{pc(0)a} \right) + \frac{i}{2} \frac{\partial}{\partial \bar{t}} g^{pc (0) <} \frac{\partial}{\partial \omega}\left(\Sigma^{a} G^{pc(0)a} \right) \right] .
\end{eqnarray}

The first order contribution to the order parameter  $\Delta_{k}^{(1)}$ is obtained from the equation:
\begin{eqnarray}
\Delta_{k}^{(1)}(\bar{t}) = \lambda \sum_{k} \int \frac{1}{2 \pi i} G^{pc(1)<} (\omega, \bar{t}) d \omega.
\label{int1}
\end{eqnarray}

The equations above provide a fully self-consistent  solution for the time dependent order parameter in the adiabatic approximation for an arbitrary time dependence of the vector potential. It can be obtained numerically, as a function of temperature and of the electric field. 

Now that we have introduced the lifetime $\tau$ of the electrons in the layer, we can discuss and obtain the adiabatic condition under which the results above are valid. 
The relevant time scales to be compared are the electronic lifetime $\tau$ and the characteristic time associated with the time variation of the vector potential ($ \propto  (\partial \mathbf{A}/\partial t)^{-1}$) in the layer.
This adiabatic condition requires that the time variation of the vector potential is much slower than the relaxation time of the electrons. Mathematically, this implies that the first order adiabatic corrections that involve the time and frequency derivatives of all quantities appearing in the expression for the lesser Green's function $G^{pc(1)<}$ are small compared to the zero order terms. Ultimately, due to the equality $\mathbf{E}=-(-1/c)\partial \mathbf{A}/\partial t$, the adiabatic condition involves a constraint on the electric field in the film and is given by, $\tilde{T} \tau/\hbar \ll 1$ where $\tilde{T}=eEv_F \tau$.     This implies that the time variation of the potential is slow compared with the characteristic relaxation time of the electrons in the layer and guarantees that the first order correction is small with respect to the zeroth order term. 

In the next section, using the first order correction for the order parameter, we calculate the phase diagram  of the dissipative superconducting film under the action of the electric field when the system has reached a steady state regime. In  this stationary regime,  the properties of the system remain unchanged in time and the time $\bar{t}$ can be replaced by the lifetime $\tau$ of the electrons in the layer ($\bar{t}=\tau=1/\Gamma$).  This is the actual time the current carriers  remain under the action of the electric  field before being scattered to  different quasi-particle states.
\subsubsection{Zero temperature phase diagram}

In order to obtain the quantum critical points and critical lines separating the normal and superconducting phases, we take  $\Delta= \Delta^{(0)}+\Delta^{(1)} = 0$ and $T = 0$. Here $\Delta$ is the actual (measurable) value of the order parameter in the presence of the electric field, under the conditions of validity of the first order adiabatic approximation. We take initially $\Delta_k^{(0)}=0$ and $T=0$. In this case, Eq.~(\ref{impo}) simplifies and we get:
\begin{eqnarray}
\label{Gpc1s}
G^{pc(1)<}\!=\!\frac{ \Delta_{k}^{(1)}}{\left[\Gamma^{2} + (\omega - \epsilon_k(\bar{t}))^{2}\right] \left[\Gamma^{2} + (\omega + \epsilon_k(\bar{t}))^{2}\right]} \left(-2\Gamma \pi [\omega \!+\! \epsilon_k(\bar{t})]^{2} \delta(\omega \!+\! \epsilon_k(\bar{t})) + \right. \nonumber\\
 \left. + 2 i \left[2 \Gamma  \omega \!+\! \pi (\omega \!-\! \epsilon_k(\bar{t}))(\omega \!+\! \epsilon_k(\bar{t}))^{2} \delta(\omega \!+\! \epsilon_k(\bar{t})) \right]\right),\nonumber\\
\end{eqnarray}
where  $\epsilon_k (\bar{t}) =\frac{1}{2m}(\mathbf{k} -e \mathbf{E} \bar{t})^2 -\frac{1}{2m}k_F^2 \approx v_F(k-k_F) -eEv_F\bar{t} \cos\theta$. Next, we replace the \textit{average} or slow time by the electronic lifetime in the layer  ($\bar{t} \rightarrow \tau$), to describe the steady state regime as we discussed before. Finally,  substituting the resulting equation in the integral (\ref{int1}) and making use of the delta functions, the remaining integral is given by,
\begin{eqnarray}
\label{delta1}
\Delta_{k}^{(1)} \!=\! \sum_{k}\frac{\lambda}{2 \pi i} \int_{0}^{\infty} \! 4 i \Gamma  \Delta_{k}^{(1)} \frac{\omega}{\left[\Gamma^{2} \!+\! (\omega \!-\! \epsilon_k(\tau))^{2}\right] \left[\Gamma^{2} \!+\! (\omega \!+\! \epsilon_k(\tau))^{2}\right]} d\omega.
\end{eqnarray}
Using that $\Delta _{k}^{(1)}=\Delta^{(1)}$ is independent of $k$, this quantity which also vanishes at the phase transition cancels out. The frequency integration can be performed to yield:
\begin{eqnarray}
\label{limite}
\frac{1}{\lambda \nu} \!=v_F\! \int_{-\pi}^{\pi}d\theta\int_{k_{F}-\delta}^{{k_{F}\!+\!\delta}} \frac{dk}{(2\pi)^{2}} \frac{1}{\epsilon_{k}(\tau)} \left[ 2\arctan \left(\frac{\epsilon_{k}(\tau)}{\Gamma}\right) \right]. 
\end{eqnarray}
We recall that  $\epsilon_k (\tau) =v_F(k-k_F) - \tilde{T} \cos\theta$ where we defined $\tilde{T}=eEv_F \tau$.

It is interesting to consider first the limit $\Gamma \rightarrow 0$ of this expression, but with the product $E \tau=E/\Gamma$ finite, when the layer decouples from the substrate.
In this limit,  performing the integrations, first in $\theta$ and then in $k$, we get:
\begin{equation}
\label{bardeen}
\frac{1}{\lambda \nu}=\frac{1}{2} \ln\left| \frac{\sqrt{1-(\tilde{T}_{c}^0/\Omega)^2}+1}{ \sqrt{1-(\tilde{T}_{c}^0/\Omega)^2}-1}\right|,
\end{equation}
where $\tilde{T}_{c}^0=eE v_F \tau$ is finite, since the electric field $E \rightarrow 0$ to keep the product $E/\Gamma=E \tau$ finite. For $(\tilde{T}_{c}^0/\Omega)\ll1$, this equation yields the following condition for the boundary of the superconducting phase:
\begin{equation}
\label{bardeen1}
\tilde{T}_{c}^0=\Delta_0.
\end{equation}
In a superconducting film the {\it depairing current density} is defined as that for which the kinetic energy of the current  carriers exceeds the binding energy of the Cooper pairs. It is then energetically favorable for the electrons to separate and cease to be superconducting. This occurs when the energy balance $\delta E=2 \Delta_0-2m v_D v_F$ becomes negative~\cite{bardeen,depairing}. The quantity $v_D$ is the drift velocity and it can be easily verified that spontaneous {\it depairing} occurs at the critical drift velocity $v_D^c= \Delta_0/m v_F$. In our case the drift velocity is given by $v_D =eE \tau/m$, such that, the condition  $v_D^c= \Delta_0/m v_F$ implies  $\tilde{T}_{c}^0=\Delta_0$ as obtained above. For current densities, such that, $\tilde{T}>\Delta_0$, the superconducting order parameter vanishes \cite{degennes}. Then, when $\Gamma \rightarrow 0$, and the film is decoupled from the substrate, it presents a {\it depairing transition} \cite{depairing} from the superconducting to the normal state for $\tilde{T}^0>\Delta_0$, i.e., for a critical electric field $E_c^0=\Delta_0/ev_F \tau$. This result relies on a correspondence, within the two fluid model, between a state with a finite normal  but  a zero superfluid one and another state with a  finite superfluid but zero normal component\cite{bardeen}.

Let us now turn on the coupling  $\Gamma$ of the layer to the substrate. To linear order in $\Gamma$ ($O(\Gamma)$) we get after performing the integrations and using that $1/  \nu = \ln(2 \Omega/\Delta_0)$,
\begin{equation}
\ln \frac{2 \Omega}{\Delta_0}=\ln \frac{2 \Omega}{\tilde{T}_c} +\frac{2}{\pi} \frac{\Gamma/\Omega}{\sqrt{1-(\tilde{T}_c/\Omega)^2}}.
\end{equation}
Neglecting terms of $O(\Gamma/\Omega.\tilde{T}_c^2/\Omega^2)$ we get,
\begin{equation}
\tilde{T}_{c}= \Delta_0 \; \exp \left({\frac{2 \Gamma}{ \pi \Omega}}\right).
\end{equation}
Then the coupling of the layer to the metallic substrate {\it increases} the critical electric field to destroy superconductivity in the layer. The physical reason for this interesting phenomenon may be that the quantity  $\Gamma$ and consequently the lifetime $\tau$ are determined  by the coupling  $t_z$ between the layer and the substrate (see the Hamiltonian, Eq.~\ref{hbl}). This coupling is responsible for the proximity effect which induces pairing in the metallic substrate reinforcing superconductivity in the whole system~\cite{mcmillan}, at least for small $t_z$.

\begin{figure}[h]
\begin{center}
\includegraphics[width=0.6\linewidth]{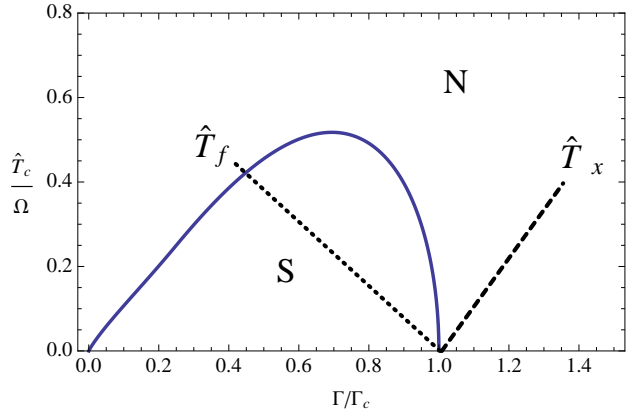}
\end{center}
\caption{(Color online) Mean-field critical line $\hat{T}_{c}(\Gamma)$  (full line) separating normal and superconducting phases, for $\lambda \nu  =0.25$. For convenience we used the variable $\hat{T}=eEv_F/\Gamma_c$  or $\hat{T}=(\Gamma/\Gamma_c)\tilde{T}$. The dotted line, $\hat{T}_f \propto |g|^{\psi}$, with $\psi=1$ for $\Gamma/\Gamma_c<1$, represents the expected shape of the critical line when fluctuations are included. The dashed line, $\hat{T}_x \propto |g|^{\nu z}$, with $\nu z=1$ is the crossover line separating the quantum critical from the quantum disordered, low field, regimes. The transition at $\Gamma=0$, $\tilde{T}_c=\Delta_0$ corresponding to the critical current in a dissipationless film (see text) appears in the variable $\hat{T}$ at $\hat{T}=0$.} 
\label{fig1}
\end{figure}

Finally, we consider Eq.~\ref{limite} in the limit that $\tilde{T} \rightarrow 0$ and for small $\Gamma$.
After integration in $k$, we are left with the following angular integral:
\begin{equation}
\frac{1}{\lambda \nu}=\frac{1}{2 \pi} \int_{- \pi}^{\pi} \ln \left[\frac{1-y^2 \cos^2(\theta)}{(\Gamma/\Omega)^2} \right] d\theta
\end{equation}
where $y=(\tilde{T}/\Omega)$.
Performing the integral, we get 
\begin{eqnarray}
\label{final}
\frac{1}{\lambda \nu} = \ln \left(\frac{\frac{1}{4}\left[1 + \sqrt{1 - (\tilde{T}_{c}/\Omega)^{2}}\right]^{2}}{\Gamma/\Omega}\right),
\end{eqnarray}
which yields,
\begin{equation}
\label{tc}
\tilde{T}_{c}=2 \Omega \left[ \left(1- \sqrt{\frac{\Gamma}{\Gamma_c}}\right) \sqrt{\frac{\Gamma}{\Gamma_c}}\right]^{1/2}
\end{equation}
for $\Gamma/\Gamma_c  \ge 1/4$. The parameter $\Gamma_c$ is given by Eq.~(\ref{gamac}) and represents the critical value of dissipation for destroying superconductivity in the layer in the absence of the electric field. Then in mean-field we find a critical line $\tilde{T}_{c}(\Gamma)$ separating the normal from the superconducting phase. 
If this line is to survive fluctuations, its non-mean-field form close to the dissipation induced quantum critical point can be obtained from a scaling approach using the properties of this QCP at $\tilde{T}_{c}=0$, $\Gamma=\Gamma_c$ determined before~\cite{livro}. We use that the electric field (or $\tilde{T}$) is a relevant perturbation at this QCP, together with standard renormalization group arguments to obtain, $\tilde{T}_{c} \propto |\Gamma-\Gamma_c|^{\psi}$, where  the shift exponent, $\psi=z/(d+z-2)$~\cite{millis,livro}. For a nearly 2D system, using $z=2$ as found before, we get $\psi=1$ (see Fig.~\ref{fig1}). The crossover line between the quantum disordered and quantum critical regime, $\tilde{T}_x \propto |\Gamma -\Gamma_c|^{\nu z}$ is also linear on the distance to the QCP, since $\nu z=1$ (see Fig.~\ref{fig1}).

Based solely on the properties of the dissipation induced QCP and the scaling expression for the electric field dependent conductivity~\cite{phillips,sondhi},
$$\sigma(E)=\xi^{2-d} F \left[ \frac{E}{|g|^{\nu(z+1)}} \right], $$
where $g=(\Gamma-\Gamma_c)/\Gamma_c$, and $\xi=|g|^{-\nu}$, we find for the conductivity of the film the scaling form,
$$\sigma(E)=\sigma_0 F \left[ \frac{E}{|g|^{3/2}} \right], $$
where we used $z=2$ and $\nu=1/2$ for the QCP at $\tilde{T}=0$, $\Gamma=\Gamma_c$ with $d_{eff}=4$.
The scaling function $F[x]$ is known in two limits~\cite{phillips}. For $x \rightarrow \infty$, $F[x]$ goes to a constant, such that the conductivity approaches a universal value at the quantum phase transition ($g=0$)~\cite{phillips}. Also for $x \rightarrow 0$, $F[x]$ has an analytic expansion, i.e., $F[x\rightarrow 0] \propto x^2 + O(x^4)$~\cite{phillips}. 
Along the critical line $\tilde{T}_c(\Gamma)$, we get,
$$\sigma(E)=\sigma_0 F \left[ \frac{E}{|g(\hat{T})|^{3/2}} \right], $$
where $g(\hat{T}_f)=0$ is the equation for the electric field dependent critical line, such that, a universal value is also reached for $E=E_c(\Gamma)$.
Because of the relation $\psi=\nu z$ between the shift and the crossover exponent~\cite{bjp}, and the equality of the scaling exponent of the conductivity in zero and finite fields, we expect  that the conductivity at this line reaches the same universal value found in Ref.~\cite{phillips} for $\Gamma=\Gamma_c$, $E=0$.

A most relevant question concerns the nature of the quantum phase transition driven by the electric field along the line $\tilde{T}_f(\Gamma)$ in Fig.~\ref{fig1}.
Since the electric field is a relevant perturbation close to the QCP at $\Gamma=\Gamma_c$, $\tilde{T}_c=0$, the critical behavior along this line is not governed by the dissipation induced QCP. Mitra et al.~\cite{mitra2} argue that this transition is in the universality class of a thermal phase transition and consequently governed by 2d thermal exponents. Indeed, in the nonequilibrium Keldysh effective action approach of this problem, when frequency goes to zero, which is the relevant limit for the critical behavior, it can be neglected with respect to the effective temperature $\tilde{T}$ associated with the electric field and the problem becomes essentially classical~\cite{mitra2}.

In figure \ref{fig1} we show the mean-field phase diagram $\tilde{T}_c(\Gamma)$ as obtained in the present work. In the limit $\Gamma \rightarrow 0$, but with the product $E . \Gamma$ finite, we recover the standard result for the critical current in a superconducting film as discussed before. The phase transition in this limit is discontinuous~\cite{degennes} and the order parameter vanishes abruptly at the critical current. An interesting possibility is that the renormalization group (RG)  flow along the critical line of the electric field-driven phase transition in the presence of dissipation is towards this point (in the variables of Fig.~\ref{fig1} this is located at $\hat{T}_c =0$, $\Gamma=0$). If this is the case, this point is a tricritical point and the exponents along the line $\hat{T}_c (\Gamma)$ should be tricritical (thermal) exponents.

 \subsection{Phase diagram for $T \not= 0 K$}

Taking $\Delta_{k}^{(0)} = 0$ and $T \not= 0 K$ equation~(\ref{impo}) becomes:
\begin{eqnarray}
G^{pc (1) <} (\omega , \bar{t})= -2 \pi\Gamma  \Delta_{k}^{(1)} \frac{f(\omega) [\omega + \epsilon_{k} (\bar{t})]^{2} \delta (\omega + \epsilon_{k})}{\left[\Gamma^{2} \!+\! (\omega \!-\! \epsilon_{k}(\bar{t}))^{2}\right] \left[\Gamma^{2} \!+\! (\omega \!+\! \epsilon_k(\bar{t}))^{2}\right]} + \nonumber\\
+4 i \Gamma  \Delta_{k}^{(1)}\frac{f(\omega) \omega \delta (\omega + \epsilon_{k})}{\left[\Gamma^{2} \!+\! (\omega \!-\! \epsilon_k(\bar{t}))^{2}\right] \left[\Gamma^{2} \!+\! (\omega \!+\! \epsilon_k(\bar{t}))^{2}\right]} + \nonumber\\
+ 2 i  \Delta_{k}^{(1)} \frac{f(\omega)[\omega - \epsilon_{k}(\bar{t})] [\omega + \epsilon_{k} (\bar{t})]^{2} \delta (\omega + \epsilon_{k})}{\left[\Gamma^{2} \!+\! (\omega \!-\! \epsilon_k(\bar{t}))^{2}\right] \left[\Gamma^{2} \!+\! (\omega \!+\! \epsilon_k(\bar{t}))^{2}\right]}.
\end{eqnarray}

\begin{figure}[h]
\begin{center}
\includegraphics[width=0.5\linewidth]{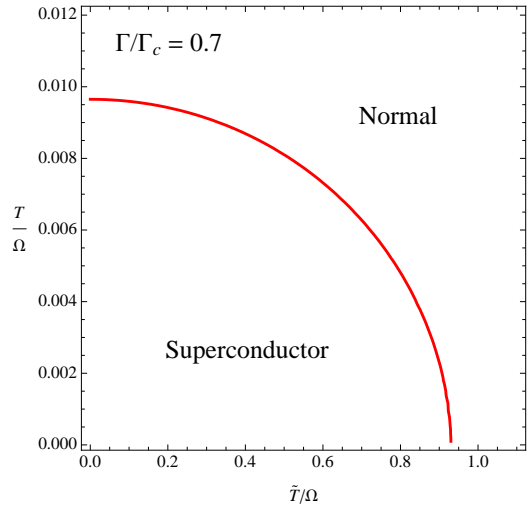}
\end{center}
\caption{(Color online) Finite temperature phase diagram $T/\Omega$   versus $\tilde{T}_{c}/\Omega$ for fixed $\Gamma/\Gamma_c=0.7$ and $\lambda \nu  =0.25$.} 
\label{fig2}
\end{figure}

Making use of the delta functions the integral (\ref{int1})) simplifies and we get,
\begin{eqnarray}
\Delta_{k}^{(1)} = \sum_{k} \frac{2 \lambda \Gamma  \Delta_{k}^{(1)}}{\pi} \int d\omega f(\omega) \frac{\omega}{\left[\Gamma^{2} \!+\! (\omega \!-\! \epsilon_k(\bar{t}))^{2}\right] \left[\Gamma^{2} \!+\! (\omega \!+\! \epsilon_k(\bar{t}))^{2}\right]}.
\end{eqnarray}
or using that $k_{F} = 2\pi\nu v_{F}/A$, 
\begin{eqnarray}
\frac{1}{\lambda \nu} = \frac{v_{F} \Gamma}{\pi^{2}} \int_{-\pi}^{+\pi} d\theta \int_{k_{F}-\delta}^{k_{F}+\delta} dk \int d\omega f(\omega)  \frac{\omega}{\left[\Gamma^{2} \!+\! (\omega \!-\! \epsilon_k(\bar{t}))^{2}\right] \left[\Gamma^{2} \!+\! (\omega \!+\! \epsilon_k(\bar{t}))^{2}\right]}.
\end{eqnarray}
From the equation above, we can obtain the phase diagram of the system at finite temperatures as a function of the field and dissipation as shown in Fig.\ref{fig2}.

\section{Conclusions}
We have generalized a Keldysh approach, previously used for time dependent impurity problems \cite{alexis,dissertacao} to treat  many-body systems close to quantum criticality. This approach allows to obtain the time dependence of the order parameter for an arbitrary time dependence of the external parameter, if an adiabatic condition is satisfied.  We have used this method to study the normal to superconductor quantum phase transition in the presence of dissipation and time dependent perturbations. Specifically, we considered a superconducting layer, under the action of a time dependent vector potential and deposited over a metallic substrate with which it can interchange electrons through a momentum non-conserving process. This is a paradigmatic problem in superconductivity that we approached from the perspective of quantum critical phenomena, as influenced by dissipation and time dependent effects. 

Initially, the dissipation induced quantum critical point was obtained in the mean field approximation yielding results in agreement with those obtained previously by Mitra~\cite{mitra}. Next, we included fluctuations close to this QCP using a Thouless criterion generalized to include dissipation.
We were able to fully characterize the dissipation induced QCP, obtaining its dynamic exponent, effective dimension and universality class. For the two-dimensional film the quantum phase transition turns out to be at the upper critical dimension and the scaling behavior in its vicinity is expected to present logarithmic contributions.

Next we treated the effect of an electric field, arising from a minimally coupled time dependent vector potential. Our approach  yields the time dependence of the order parameter in the ordered phase. We studied the phase diagram in the presence of the electric field and dissipation in the nonequilibrium stationary state. In the neighborhood of the dissipation induced QCP ($g \rightarrow 0$), and also for small electric fields ($E << |g|^{3/2}$), several properties can be obtained  using general scaling results and the properties of this QCP, in agreement with previous results.
Within the BCS and adiabatic approximations, we have obtained the full electric field versus temperature, versus dissipation phase diagram. In particular at $T=0$ and as $\Gamma \rightarrow 0$ we find a pair-breaking transition associated with a critical current.

\subsection*{Acknowledgments}

This work was supported in part by CNPq - Conselho Nacional de
Desenvolvimento Cient\'ifico e Tecnol\'ogico (Brazil) and CAPES - Coordena\c{c}\~ao de Aperfei\c{c}oamento de Pessoal de n\'ivel Superior (Brazil).  M. A. C. also
thanks FAPERJ - Funda\c{c}\~ao de Amparo \`a Pesquisa do Estado do Rio
de Janeiro for partial financial support. The authors would like to thank Alexis Hernandez, Claudine Lacroix,  Denis Feinberg and  Regis Melin  for comments and interesting discussions.


\begin{thebibliography}{100}

\bibitem{ventra} M. di Ventra, \textit{Electrical Transport en Nanoscale Systems} (Cambridge University Press, Cambridge, 2008); S. Datta, \textit{Electronic Transport in Mesoscopic Systems} (Cambridge University Press, Cambridge, 1995).


\bibitem{hertz} J. A. Hertz, Phys. Rev. B {\bf14}, 1165 (1976).

\bibitem{chakra} Philipp Werner, Klaus V\"oker, Matthias Troyer, and Sudip Chakravarty, Phys. Rev. Lett. \textbf{94}, 047201 (2005).


\bibitem{tyablikov} S. V. Tyablikov, \textit{Methods in the Quantum Theory of Magnetism}, (Plenum, New York, 1967).

\bibitem{kadanoff} L. P. Kadanoff and G. Baym, \textit{ Quantum Statistical Mechanics} (Benjamin, New York, 1962).

\bibitem{keldysh} L. V. Keldysh, Sov. Phys. JETP \textbf{20}, 1018 (1965).

\bibitem{phillips} Denis Dalidovich and Philip Phillips, Phys. Rev. Lett. \textbf{93}, 027004 (2004);
D. Dalidovich and P. Phillips, Phys. Rev. Lett. 84, 737 (2000).

\bibitem{sondhi} A. G. Green and S. L. Sondhi, Phys. Rev. Lett. \textbf{95}, 267001 (2005); see also M. di Ventra, \textit{Electrical Transport en Nanoscale Systems} (Cambridge University Press, Cambridge, 2008).


\bibitem{mitra} A. Mitra, Phys. Rev. B, \textbf{78}, 214512 (2008).

\bibitem{takei} S. Takei and Y.B. Kim, Phys. Rev. \textbf{B78},165401 (2008).

\bibitem{hogan} P. M. Hogan and A. G. Grenn, Phys. Rev. \textbf{B78}, 196104 (2008).

\bibitem{feldman}  D. E. Feldman, Phys. Rev. Lett. \textbf{95}, 177201 (2005).

\bibitem{mcmillan} W. L. McMillan, Phys. Rev. \textbf{175}, 537 (1968); Ya. V. Fominov, N. M. Chtchelkatchev and A. A. Golubov, Phys. Rev. \textbf{B66}, 014507 (2002).

\bibitem{alexis} A. R. Hernandez, F. A. Pinheiro, C. H. Lewenkopf e E. R. Mucciolo, Phys. Rev. B, \textbf{80}, 115311 (2009).

\bibitem{dissertacao} Fernanda Deus, Alexis R. Hernandez and Mucio A. Continentino, J. Phys.: Condens. Matter \textbf{24},    356001   (2012).


\bibitem{bardeen} John Bardeen, Rev. Mod. Phys., \textbf{34}, 667 (1962).

\bibitem{depairing} see D. Dew-Hughes, Low Temp. Phys. \textbf{27}, 713 (2001).

\bibitem{depairing2} A. Yu. Rusanov, M. B. S. Hesselberth, and J. Aarts, Phys. Rev. {\bf B70}, 024510 (2004).


\bibitem{yu} M. Yu. Kupriyanov and V. F. Lukichev, Fiz. Nizk. Temp. {\bf 6}, 445
(1980) [Sov. J. Low Temp. Phys. {\bf 6}, 210 (1980)].

\bibitem{degennes} P. G. de Gennes, \textit{Superconductivity of Metals and Alloys}, p.182, Perseus Books, Massachussets, USA. 1999.

\bibitem{maki} K. Maki, Progr. Theor. Phys. \textbf{29}, 333 (1963); ibid \textbf{31}, 731 (1964).K. Maki, in Superconductivity, edited by R. D. Parks (Dekker,
New York, 1969), Vol. 2.

\bibitem{livro} \textit{Quantum scaling in many-body systems}, M.~A.~Continentino, World Scientific, Singapore (2001).


\bibitem{aproxbcs} J. Bardeen, L. N. Cooper e J. R. Schrieffer, Phys. Rev. \textbf{106}, 162 (1957).

\bibitem{aline} Aline Ramires and Mucio A. Continentino,  J. Phys.: Condens. Matter {\bf 22} 485701 (2010).

\bibitem{Thouless} D. J. Thouless, \textit{Annals of Physics} \textbf{10},
553 (1960). 

\bibitem{langreth} D. C. Langreth, \textit{Linear and Nonlinear Electron Transport in Solids} (Plenum, New York, 1976).

\bibitem{millis} A. J. Millis, Phys. Rev. B 48, 7183 (1993).

\bibitem{mitra2} Aditi Mitra, So Takei, Yong Baek Kim and A. J. Millis, Phys. Rev. Lett. \textbf{97}, 236808 (2006);Aditi Mitra and Andrew J. Millis, Phys. Rev. \textbf{B77}, 220404 (2008); 

\bibitem{bjp} M. A. Continentino, Brazilian Journal of Physics, \textbf{41}, 201 (2011).


\end{thebibliography}
\end{document}